\documentclass[aps,showpacs,pre,twocolumn]{revtex4}
\usepackage{epsfig}
\usepackage{amsmath}
\usepackage{amssymb}
\usepackage{graphicx,natbib}%amssymb
\usepackage{setspace}
\begin{document}
%\doublespacing

%\bibliographystyle{svjour}%apsrev}%prsty}

\title{Growing Directed Networks: Stationary in-degree probability for arbitrary out-degree one}
\author{Daniel Fraiman}
\affiliation{%
%\institute{
Departamento de Matem\'atica y Ciencias, Universidad
de San Andr\'es, Buenos Aires, Argentina.}

%\abstract{
\begin{abstract}
%\abstract{
We compute the stationary in-degree probability,
$P_{in}(k)$, for a growing network model with directed edges and
arbitrary out-degree probability. In particular, under
preferential linking, we find that if the nodes have a light tail
(finite variance) out-degree distribution, then the corresponding
in-degree one behaves as $k^{-3}$. Moreover, for an out-degree
distribution with a scale invariant tail, $P_{out}(k)\sim
k^{-\alpha}$, the corresponding in-degree distribution has
exactly the same asymptotic behavior only if $2<\alpha<3$
(infinite variance). Similar results are obtained when
attractiveness is included. We also present some results on
descriptive statistics measures
%descriptive statistics
 such as the correlation between the number of in-going
links, $D_{in}$, and outgoing links, $D_{out}$, and the
conditional expectation of $D_{in}$ given $D_{out}$,
 and we calculate these measures for the WWW
network. Finally, we present an application to the scientific
publications network. The results presented here can explain the
tail behavior of in/out-degree distribution observed in many real
networks.%}
\end{abstract}
%\PACS{05.65.+b, 89.75.Kd, 87.23.Ge, 02.50.Cw}
 \maketitle

\section{Introduction}

Barab\'asi and Albert~\cite{net:bara_science} discovered that
several networks in nature have a strange topological
characteristic: they have a
scale-free~\cite{net:review1,net:review2,net:jeong} degree
distribution, $P(k)\sim k^{-\alpha}$, where the degree of a
vertex is defined as the total number of its connections.
Nowadays, this empirical behavior is confirmed in a great number
of completely different empirical networks, from biological
networks to e-mail networks, including scientific publication
networks. In~\cite{net:bara_science} they  also proposed a model
(B-A model) for explaining this behavior. The model can be
formulated as follows: 1) start with a network with $N$ nodes,
connected by $j$ edges in an
 arbitrary way, and 2) at each time step a new node, with $m$
  edges, appears, and each of edges connects to
the existing nodes according to some probability law, $\pi$. The
probability that a new edge attaches to a node with degree $k$,
$\pi^k$, was defined~\cite{net:bara_science} as proportional to
the degree of the node. In particular, they showed that with this
attachment law,
\begin{equation}
\pi^k=\frac{k N^k}{\underset{j \in \mathbb{N}}{\sum}jN^j},
\end{equation}
where $N^k$ is the number of nodes with degree $k$, the
stationary degree distribution has a power law tail, $P(k)\sim
k^{-3}$.
 In~\cite{net:mendes} they computed the
stationary degree probability (not only the tail behavior) or
limit degree distribution for a model similar to the B-A one, but
for a generalization of the preferential linking attachment law.
They introduced a new parameter, the attractiveness, $A$ (in their case $A \geq
0$), and defined the attachment law as:
\begin{equation}\label{mendes}
\pi^k = \frac{(A+k)N_{in}^k}{\underset{j \in \mathbb{N}}{\sum}
(A+j)N_{in}^j},
\end{equation}
where $N_{in}^k$ is the number of nodes with in-degree equal $k$.
They found in this case that $P(k)\sim k^{-(2+A/m)}$,
 being more flexible for comparing to empirical networks.
 Typically,  degree distribution of real networks satisfy,
 $P(k)\sim k^{-\alpha}$ with $2\leq \alpha\leq 3$.
But the B-A model and similar ones~\cite{net:mendes}, no matter
which is the attachment law, have a mayor drawback, the number
($m$) of edges that arise from new nodes is a fixed number. In
almost all real networks, the new nodes do not have the same
number of edges. On the other hand, the
 number of edges of a random selected new node (from a real network) is
 a random variable.
So, in order to be more realistic, we will study the behavior
of the B-A model
 when new nodes with a random number of edges appear, but in the more general
  context of directed growing networks. In this context new questions arises.

Directed networks are characterized by the fact that the edges are
directed (arrows), each node has edges that point at it, and
others that born in it. The in-degree of a node is defined as the
number of incoming edges, and the out-degree as the number of its
outgoing edges.
 The most studied directed growing networks
have been the WWW network~\cite{net:kapri,net:tadic,serrano}, and
the scientific publications network~\cite{net:redner}. In the
first one, each node represents a web page and the hyper-links
(references to other web pages) represents the directed edges or
links. In the second one, each paper is a node, and its
references the directed links.  In this last case, the in-degree
distribution represents the distribution of citations for a
random selected paper,
 and the out-degree distribution represents the number of references of a random selected paper.
Empirical directed growing networks follow in general one of two possible behaviors.
 In the first case they have an out-degree
exponential distribution, $P_{out}(k) \sim a^k$ ($0<a<1$), or an
out-degree distribution taking finitely many values, associated
with an in-degree one distribution with a power law tail
$P_{in}(k) \sim k^{-\alpha}$ where typically $\alpha\approx 3$.
In the second case the out-degree distribution satisfies
$P_{out}(k)\sim k^{-\beta}$, and is associated with $P_{in}(k)\sim
k^{-\alpha}$ with $\alpha \approx \beta$. Examples, such as
biological, WWW,  or communication
 networks, can be found
in~\cite{net:review1,net:review2,net:review3,net:jeong}.

In this paper, we address the question of why the empirical
growing directed networks show this strange general behavior for
the tail of the in/out degree distributions. We study a particular
growing network model (a generalization
of~\cite{net:bara_science} to be precise), obtaining the
stationary joint in-out degree distribution, $P_{in,out}(j,k)$,
and some of its derivatives, such as the marginal distribution,
$P_{in}(k)$, the covariance, and the conditional expectation of
the number of in-links given the number of out-links. In
particular, studying in detail $P_{in}(k)$, we prove (for the
model presented here) that it is expected to observe the in/out
tail behavior reported for real
networks~\cite{net:review1,net:review2,net:jeong}. Finally we
present an application to the most ``pure'' (extremely few double
arrows) growing directed network: the scientific publication
network. In this application, we show the relevance of having an
expression for the limit in-degree distribution ($P_{in}(k)$) for
an arbitrary out-degree one ($P_{out}(k)$).

%the show the relevance of obtaining the limit in-degree
%distribution ($P_{in}(k)$) for an arbitrary out-degree one
%($P_{out}(k)$).

%\section{GROWING DIRECTED MODEL}

\section{Growing Directed Network Model}
 Before describing the model, it is important to remark that real
directed growing networks have in general a considerable asymmetry
between the in-links and out-links of a node. For example, nobody
will care much about how many references (out-links) an own paper
has, but people are interested in the number of cites (in-links)
that their own paper has. That is why we are going to treat the
out-links from a new node and the in-links in a
 completely different way. In
particular, a node can receive (with positive probability), a
connection from a new node at any moment, but typically a node can
not change who their pointers (the set of nodes it is pointing to)
are.
 This is very clear in
the scientific publications network. In this network the
in-degree distribution has been extensively
study~\cite{net:redner,net:tadic}, whereas the out-degree
distribution has been poorly
reported~\cite{net:price,net:lambio}. Nevertheless, in the case
of the WWW network, the outgoing links (hyper-links) can change
at any moment and new hyper-links can be aggregated or old
hyper-links can be redirected. In~\cite{net:kapri,net:tadic} they
proposed some models for describing this network taking into
account the characteristics mentioned above. However these models
do not consider that the new nodes have a particular out-degree
distribution, i.e.  the models are constructed under the
hypothesis that new nodes have a fixed number of out-links. The
mayor problem of both models is that the nodes (webpages) do not
have a controlled number of out-links, they can have a huge
number of them which does not seem realistic. Our strategy for
modeling these networks is completely different to the ones
proposed in~\cite{net:kapri,net:tadic}, for us, most of the
variability in the number of out-links is explained when the node
appear, defined as ``intrinsic'' variability,  and not as a
product of updating nodes. We think that in many real networks
the updating of nodes can give a small correction compared with
the ``intrinsic'' variability. This assumption is at the core of
our model. In a real network the ``intrinsic'' variability is
given by different reasons that are hard to know (why does a
randomly selected scientific paper has a number
 of references with some particular distribution?), but typically
 the problem of trying
to understand it is not a mayor question.

 Now, we describe the growing network model: 1) initially the network consists of $N$
 nodes connected in a given arbitrary way, 2) at each time step, say time step $n+1$, a node
 with $D_{out}$ outgoing-edges appear, where $D_{out}$ is a random
 variable ($\underset{k\in \mathbb{N}}\sum P(D_{out}=k)=1$), and 3) each new directed edge points out to an existing node
with some probability law $ \pi_{n+1}$ (uniform, preferential
linking, etc.). Fig.~\ref{model} shows an scheme of the model.
\begin{figure}[ht]
\begin{center}
\includegraphics[height=0.1\textwidth]{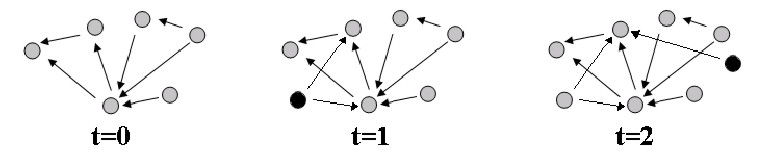}
\end{center}\label{model}
\caption{Scheme of the growing network model. In each temporal
step a new node (shown in black) with $D_{out}$ out-links appear;
these links point towards  existing nodes.  $D_{out}$ is not a
fixed number, on the contrary is a random variable. The degree
vector at time 0, and 1 is:
$\vec{N}_{0}=(1,4,0,0,1,0,0,0,...,0,...)$,
$\vec{N}_{1}=(1,4,1,0,0,1,0,0,...,0,...)$. }
 \label{model}
\end{figure}
 If $\pi_{n+1}$ is an arbitrary
function that depends on the degree vector at time $n$,
$\vec{N}_{n}=(N^1_{n},N^2_{n},...,N^{k}_{n},...)$
 and/or $\vec{N}_{in,n}$ ($\vec{N}_{out,n}$), then the growing
network model, described above is a Markov chain taking values in
$\mathbb{N}_0^{\mathbb{N}}$ or $\mathbb{N}_0\times
\mathbb{N}_0^{\mathbb{N}^2}$
 with transition probabilities given by $\pi_{n+1}$. In this work (under the Markovian hypothesis),
 we show an easy way to compute stationary (in/out) degree probabilities for arbitrary
$\pi_{n+1}$. An important part of this article is devoted to the
study of the model under the law:
 \begin{equation}\label{law1}
% \begin{aligned}
\pi_{n+1}^k := \frac{(A+k)N_{n}^k}{\underset{j \in
\mathbb{N}}{\sum} (A+j)N_{n}^j},
 \end{equation}
and in Section 2.4 we show some results under different $\pi$'s.
The law of eq.~\ref{law1} corresponds to preferential linking on
degree with attractiveness. This probability is well defined for
values of $A$ greater or equal to -$B$, where
 \begin{equation}\label{b}
B=\underset{k}{min}\{k: P(D_{out}=k)>0\}.
 \end{equation}
For this attachment law, the model is in fact an extension of the
Albert-Barab\'asi model,  although in this case $D_{out}$ is a
random variable with an arbitrary distribution, $P(D_{out}=k)$
with $k\in \mathbb{N}$, and the edges are directed. The limit
(stationary) in-degree distribution and the limit degree
distribution have not been reported, even for simple cases as
$D_{out}$ taking values 1 and 2, with probabilities $p_1$ and
$1-p_1$ respectively.
 Moreover, even in the undirected case, it is not known if in general the
limit degree distribution ($P(k)$) satisfies a superposition
principle (linear combination).

\subsection{Stationary Probabilities}
The number of out-links does not depend on time (see Appendix A
for additional details), therefore, the limit out-degree
distribution satisfies $P_{out}(k)\equiv P(D_{out}=k)$.
 Note that the out-degree distribution is defined a priori (in accordance with the specific network),
 imposing in this way the asymmetry mentioned before between the in and out
 links.
  We are
interested in obtaining the limit degree distribution, $P(k)$,
and the limit in-degree one, $P_{in}(k)$. In order to compute
this last probability distribution, we first compute the
stationary joint degree and out-degree distribution,
$P_{deg,out}(j,k)$.
 If the
network is distributed according to the stationary probability,
then the probability that a randomly chosen node has $k$ out-links and $j$ total links,
 $\vec{D}=(D,D_{out})=(j,k)$, is given by:
\begin{equation*}
 P_{deg,out}(j,k)=P(\vec{D}=(j,k))=\underset{n \rightarrow \infty}{\lim}
\frac{N^{j,k}_{deg,out,n}}{\underset{j,k \in
\mathbb{N},\mathbb{N}_o}{\sum}{N^{j,k}_{deg,out,n}}}
\end{equation*}
where $N^{h,i}_{deg,out,n}$ is the number of nodes with $h$ total
links from which $i$ are out-links at time $n$. The last equality
holds by the Law of Large Numbers.
 Clearly, the joint in-out degree can be
computed from this last one, $P_{in,out}(j-k,k)=P_{deg,out}(j,k)$, and also the in-degree
and degree probability taking marginal distributions.

$N^{j,k}_{deg,out,n+1}$ depends on: 1) $N^{j,k}_{deg,out,n}$, and
2) the transition probabilities, $\pi_{deg,out,n+1}$. As it is
usual for Markov chains, we associate to the transition
probabilities of this chain some random variables that we now
describe. In the first place, there is the out-degree, $D_{out}$,
of the new node. Secondly, we consider at each time $n+1$ a
sequence of independent and identical distributed bivariate
random vectors
$\{\vec{Z}_i\}$, %\{\vec{Z}_i\}_{1\leq i \leq n}$,
 taking value $(j,k)$, $j,k\in \mathbb{N}$, with probability $\pi_{deg,out,n+1}^{j,k}$, which
depends on the state of the chain at time $n$.
 This way, the growing network
dynamics can be written as:
%\begin{widetext}
\begin{equation}\label{din}
%\begin{aligned}
N_{deg,out,n+1}^{j,k}=N_{deg,out,n}^{j,k}+\Delta_n^{j,k} \quad \forall j\geq k \in \mathbb{N}\\
%,\quad \forall j>k \in N \\
%& X_{deg,out,n+1}^{j,j}=X_{deg,out,n}^{j,j}+\Delta_n^{j,j}, \\
% \end{aligned}
\end{equation}
%\end{widetext}
where
\begin{equation}\label{deta}
\Delta_n^{j,k}=
\begin{cases}
&\overset{D_{out}}{\underset{i=1}{\sum}}\delta_{\vec{Z}_i=(j-1,k)}-\delta_{\vec{Z}_i=(j,k)}  \quad \mbox{for $j>k$} \\
&\delta_{D_{out}=j}-\overset{D_{out}}{\underset{i=1}{\sum}}\delta_{\vec{Z}_i=(j,j)}\quad \quad \ \ \mbox{for $j=k$}\\
 \end{cases}
\end{equation}
The random vector $\vec{Z}_{i}$ indicates to which type of node
the $i$ link (of the new node) is pointing to. For example, if
$\vec{Z}_{1}=(3,2)$, a new link is pointing to an existing node
with 2 out-links and 1 in-link (or 3 total links). Clearly, in
order to have a good representation of the growing network
process,  the probability law of $Z_{i}$ must be equal to
$\pi_{deg,out,n+1}^{j,k}$, as we impose. Equations~\ref{din}
and~\ref{deta} can be read in the following way: if at time $n+1$
a new node with $D_{out}=m$ out-links is aggregated, then
$N_{deg,out,n+1}^{m,m}$ grows by one, and $m$ components of the
degree vector undergo a ``shift''. %of the kind
%$\vec{N}_{deg,n+1}=(..,X^k_{deg,n}-1,X^{k+1}_{deg,n}+1,X^{k+2}_{deg,n},...)$.
As the network continues to grow, the goal is to find whether
there exists a limit distribution for the in-out degree. For very
large values of $n$, given a randomly selected node, what is the
probability that this one has $j$ links, of which $k$ are
out-links, $P_{deg,out}(j,k)$?.

The following property shows a way of computing

$P_{deg,out}(j,k)$ which has interest on
itself.

\textbf{Property:}  $P_{deg,out}(j,k)$ is the solution of:
\begin{equation}\label{proper}
P_{deg,out}(j,k)=\langle \Delta_n^{j,k}/\Theta_n \rangle \quad
\forall j\geq k \in \mathbb{N},
\end{equation}
where $\Theta_n$ is the event that imposes that the empirical
distribution at time $n$ is equal to the stationary distribution,
i.e.  $\Theta_n=\{\frac{N^{h,i}_{deg,out,n}}{\underset{l,m \in
\mathbb{N}}{\sum}N^{l,m}_{deg,out,n}}=P_{deg,out}(h,i) \quad
\forall h,i \in \mathbb{N}\}$. The preceding property says that
if the process at time $n$ is distributed according to the
stationary probability, $P_{deg,out}$, it will remain
there in expectation. %Note that the  $\vec{Z}_n \equiv
%\frac{\vec{X}_{n}}{\underset{k \in N}{\sum}X_{n}^k}$ process is
%not a Martingale one.
This technique for finding stationary probabilities seems much
easier (see Appendix B) than previous
approaches~\cite{net:mendes,net:bara_science,net:tipos_attach}.

Using the property mentioned above and eq.~\ref{deta}, it is easy
to see that the stationary joint deg-out distribution,
$P_{deg,out}$, satisfies:
\begin{equation}\label{din_gral2}
\begin{aligned}
& P_{deg,out}(j,k)=\pi_{deg,out}^{j-1,k}\langle D_{out}\rangle-\pi_{deg,out}^{j,k}\langle D_{out}\rangle\\
& P_{deg,out}(j,j)=P_{out}(j)-\pi_{deg,out}^{j,j}\langle D_{out}\rangle\\
\end{aligned}
\end{equation}
for $j > k \in \mathbb{N}$, where $\langle
D_{out}\rangle=\overset{\infty}{\underset{k=0}{\sum}}kP_{out}(k)$.
These two equations contain all the information about the limit
joint in-out degree distribution, being a crucial result in this
paper. It is important to note that since we have conditioned on
the fact that at time $n$ the process is distributed according to
the stationary probability, the link attachment probability does
not depend on time. Now, $\pi_{deg,out}^{j,k}$ denotes the
stationary probability that a new link (from a new node) point to
an existing node with $j-k$ in-degree links and $k$ out-degree
links. Under preferential linking on degree with attractiveness
(eq.~\ref{law1}), the stationary attachment law remains:
\begin{equation}\label{pref2}
\pi_{deg,out}^{j,k}=\frac{j+A}{\langle D \rangle+A}P_{deg,out}(j,k).
\end{equation}
where $\langle D
\rangle=\overset{\infty}{\underset{k=1}{\sum}}kP_{deg}(k)$.
 The marginal distribution of eq.~\ref{pref2},
 $\pi^{k}=\overset{k}{\underset{j=1}{\sum}}\pi_{deg,out}^{k,j}$,
 is the stationary version of $\pi^{k}_{n+1}$ presented in
 eq.~\ref{law1}.
 Replacing eq.~\ref{pref2} in eq.~\ref{din_gral2}, and
 using $\langle D \rangle=2\langle D_{out} \rangle$ (for each new node with $k$ out-links,
  the total degree increases by $2k$) we obtain:
\begin{equation}\label{pref3}
 P_{deg,out}(j,k)=\frac{\Psi(j+A,3+\delta)}{\Psi(k+A,2+\delta)}P_{out}(k),
\end{equation}
where $\Psi(a,b)\equiv
\frac{\Gamma[a]\Gamma[b]}{\Gamma[a+b]}=\int^1_0t^{a-1}(1-t)^{b-1}dt$
(Beta function), and $\delta=A/\langle D_{out}\rangle$. From
eq.~\ref{pref3}, taking marginal distributions is trivial to
obtain:
\begin{equation}\label{pref4}
\begin{aligned}
&  P_{in,out}(j,k)=\frac{\Psi(j+k+A,3+\delta)}{\Psi(k+A,2+\delta)}P_{out}(k) \quad \quad (a)\\
&P(k)=\Psi(k+A,3+\delta)
\overset{k}{\underset{j=1}{\sum}}\frac{P_{out}(j)}{\Psi(j+A,2+\delta)} \quad \quad (b)\\
&P_{in}(k)=\overset{\infty}{\underset{j=1}{\sum}}P_{out}(j)\frac{\Psi(j+k+A,3+\delta)}{\Psi(j+A,2+\delta)}  \quad \quad \quad (c).\\
\end{aligned}
\end{equation}
Eq.~\ref{pref4} shows the joint stationary in-out degree probability,
 the degree distribution and the in-degree distribution.
 In the stationary regime (for the probability) the proportion
of nodes with $j$ in-links and $k$ out-links (eq.~\ref{pref4}
(a)), depends on the attractiveness, and on the out-degree
distribution through two quantities: $\langle D_{out}\rangle$ and
$P_{out}(k)$. The same happens for $P(k)$ and $P_{in}(k)$.
Eq.~\ref{pref4} (b) shows the stationary degree probability for
arbitrary out-degree distribution (see Appendix B for a simpler
derivation). Note that just by replacing $P_{out}(k)$ by
$\delta_{k=m}$ (this means a non-random $D_{out}$ and equal to
$m$) we obtain the known result~\cite{net:mendes} for undirected
networks. Eq.~\ref{pref4} (c) constitutes one of the main results
of the paper.
 Replacing $P_{out}(k)$ by the empirical
value, we can check whether the model is adequate for the network
under study.
 Moreover, it is possible to see that a superposition
 principle does not hold, either for $P(k)$, $P_{in}(k)$, or
 $P_{in,out}(k,j)$.
  They cannot be written as
 $P(k)=\overset{\infty}{\underset{j=1}{\sum}}P_{out}(j)Q_j(k)$, where $Q_j(k)$ is the stationary
 probability for a fixed number $j$ of out-links. The superposition principle
 will be valid for the three limit distributions only when
 the attractiveness vanishes (preferential linking).
 In this way, the preferential linking generalization (the inclusion of attractiveness)
 introduced in~\cite{net:mendes} has the advantage of enlarging the
 power exponent values of the degree distribution, with the drawback of loosing a superposition principle.
 If we allow the appearance of new nodes with
 zero out-links ($P(D_{out}=k)=P_{out}(k)$ with $k\in \mathbb{N}_o$), then
 the results presented in equations~\ref{pref4} (b) and (c), still
 hold after switching the initial index in the summation from 1 to 0 and taking $k\in \mathbb{N}_o=\mathbb{N}\cup\{0\}$.
 In this last case, the attractiveness must be greater o equal zero (see eq.~\ref{b}).

%\subsection{Some informal checks}
%\subsection{Some informal checks before modeling}
%\subsection{From data to model}                % juan tessi 11-319-16158
\subsection{Descriptive Statistics}
Before trying to describe a real network by a model, some first
checks are recommendable.  One typical measure that has been
extensively used is the clustering coefficient, that is a measure
of how connected the neighbors of a node are. We are going to
discuss much simpler descriptive measures that also serve as
tools for looking for the ``best'' model. Therefore, it is
important to have analytical devices for comparing with real data
in the search of a good model.

\subsubsection{Covariance and conditional expectation}
A measure of dependence between the in-degree and the out-degree
 can give an idea of which is the attachment law that
  better describes the empirical data. The covariance between $D_{out}$ and $D_{in}$,
$Cov(D_{in},D_{out})= \langle D_{in}D_{out} \rangle-\langle
D_{in}\rangle \langle D_{out}\rangle$ is an adequate statistical
measure for this purpose. For example, in the case where the law
of attachment is preferential linking on in-degree
(eq.~\ref{mendes}) this measure is obviously zero. For the case
studied in detail here, preferential linking on degree
(eq.~\ref{law1}), it is straightforward to see that the
covariance between $D_{out}$ and $D_{in}$ in the particular case
$A=0$, satisfies the following equation:
\begin{equation}\label{cov}
Cov(D_{in},D_{out})=\frac{1}{2}Cov(D,D_{out})=Var(D_{out})
\end{equation}
where $Var(D_{out})=Cov(D_{out},D_{out})$. The covariance is
always positive or zero (for non random $D_{out}$), as it is
expected for this type of attachment law. Eq.~\ref{cov}
instead can be written in terms of the correlation, $r=\frac{Cov(D_{in},D_{out})}{\sqrt{Var(D_{in})Var(D_{out})}}$,
 in the following way:
  \begin{equation}\label{cor}
r=\sqrt{\frac{Var(D_{out})}{Var(D_{in})}}.
\end{equation}
It is surprising that the correlation satisfy this simple
relation between the standard deviations, r is the ratio between
$\sigma_{out}$ ($\sqrt{Var( D_{out})}$) and $\sigma_{in}$
($\sqrt{Var(D_{in})}$). Since the correlation coefficient is
always less or equal 1, we obtain the following inequality:
    \begin{equation}\label{cor2}
Var(D_{out})\leq Var(D_{in}).
\end{equation}
Although it is very easy for real network to estimate the
variance of the number of out and in links, and also the
covariance (or correlation) between the in and out-degree, these
measures are not typically reported (see Appendix C for results
on the WWW network).

On the other hand, the first right term of the covariance always
satisfies:
\begin{equation}\label{aux}
\langle D_{in}D_{out}\rangle=\underset{k \in
\mathbb{N}}{\sum}k\langle D_{in}/D_{out}=k \rangle P_{out}(k),
\end{equation}
where $\langle D_{in}/D_{out}=k \rangle$ is the conditional
expectation of  the number of in-links given that the node has $k$
out-links. From equations~\ref{cov} and~\ref{aux} it is very easy
to see that:
\begin{equation}\label{espcond}
 \langle D_{in} /D_{out} \rangle= \frac{1}{2}\langle D /D_{out} \rangle=D_{out}.
\end{equation}
The relationship between $\langle D_{in} /D_{out} \rangle$ and
$D_{out}$  can be a second check to make before modeling. For a
real network this can be done in the following way, choose all
the nodes that have a number $D_{out}$ of outgoing links, and
take the mean of the number of in-links over this set of nodes.
If the conditional mean is equal to $D_{out}$ for all values of
$D_{out}$, then this is an indication that the model can be
adequate.

\begin{figure}
\begin{center}
\includegraphics[height=0.8\textwidth]{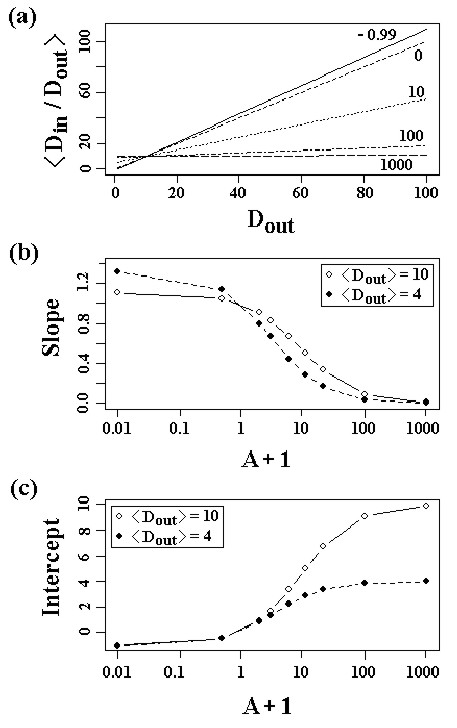}%{fig2_new.eps}
\end{center}
\caption{(a) Conditional expectation of in-degree given the
out-degree. Each straight line correspond to a different value of
attractiveness (specified in the graph). (b) Slope and (c)
Intercept of the type of straight lines  shown in (a) as a
function of  the attractiveness for two different values
of $<D_{out}>$. }\label{esperanza}% decia(3.825) (1-0.3)/(1-po)=0.7622781
\end{figure}

For non null attractiveness it is hard to obtain analytical
results, nevertheless, we compute numerically $\langle D_{in}
/D_{out} \rangle$ for different values of $D_{out}$ and
attractiveness. From eq.~\ref{pref4} (a) and the definition of
conditional expectation, it is easy to obtain:
\begin{equation} \label{espec}
\langle D_{in} /D_{out} \rangle=\underset{j \in \mathbb{N}}{\sum}j\frac{\Psi(j+D_{out}+A,3+\delta)}{\Psi(D_{out}+A,2+\delta)}.
%=\underset{j \in \mathbb{N}}{\sum}j\frac{P_{in,out}(j,D_{out})}{P_{out}(D_{out})}
\end{equation}
Fig.~\ref{esperanza} (a) shows the numerical results of $\langle
D_{in} /D_{out} \rangle$ based on eq.~\ref{espec}.
 For any value of the attractiveness and $\langle D_{out}\rangle$,
 the conditional expectation follows a linear relation with $D_{out}$:
\begin{equation}\label{espcond2}
 \langle D_{in} /D_{out} \rangle= f(A, \langle D_{out}\rangle)D_{out}+g(A,\langle D_{out}\rangle).
\end{equation}
 The slope, $f(A, \langle D_{out}\rangle)$,  and the intercept, $g(A,\langle D_{out}\rangle)$,
  of this straight line satisfies:
\begin{equation}
\begin{aligned}
& \underset{A\rightarrow \infty }{lim}f(A, \langle D_{out}\rangle)=0 \\
& \underset{A\rightarrow \infty }{lim}g(A, \langle D_{out}\rangle)=\langle D_{out}\rangle, \\
\end{aligned}
\end{equation}
as it is shown in Fig.~\ref{esperanza} (b) and (c).   For
positive values of attractiveness  the slope is smaller than one,
going to zero as the attractiveness goes to infinity. In the case
$A\rightarrow \infty$, $D_{in}$ and $D_{out}$ are independent
(always with the same expectation).
 Finally, for negative values of $A$ the slope is greater than one.
Studying the empirical relationship between $\langle D_{in}
/D_{out} \rangle$ and $D_{out}$ can give some insight on the
model. Moreover, if this relationship is linear,  from
Fig.~\ref{esperanza} (b) and (c),
 it is possible to have a first estimation of %it is also possible to estimate
 the attractiveness. In Appendix C we show the statistical
 measures presented here for the WWW network.

  It is important to note that equations~\ref{cov}
  (which includes~\ref{cor},~\ref{cor2}), and~\ref{espcond2}
  (which include~\ref{espcond}) holds for any out-degree distribution
($P_{out}(k)$). These results do not depend on the details (shape)
of the out-degree distribution.
 Nevertheless, there exist some measures that do not share this nice property. For example,
 the conditional number of out-links given the number of in-links, $\langle D_{out} /D_{in} \rangle$,
  depends explicitly on $P_{out}(k)$, as can be seen in the following equation:
  \begin{equation}\label{espcond3}
 \langle D_{out} /D_{in}=k \rangle= \frac{\overset{\infty}{\underset{j=1}{\sum}}j\frac{\Psi(k+j+A,3+\delta)}{\Psi(j+A,2+\delta)}P_{out}(j)}{\overset{\infty}{\underset{h=1}{\sum}}\frac{\Psi(h+k+A,3+\delta)}{\Psi(h+A,2+\delta)}P_{out}(h)}.
\end{equation}

  Next, we present another measure useful for model selection.

\subsubsection{Relationship between the distribution tails}
\begin{figure}
\begin{center}
\includegraphics[height=0.4\textwidth]{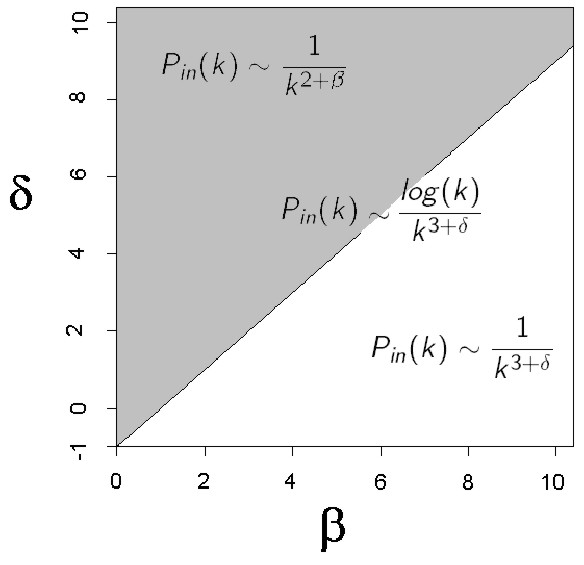}
\end{center}
\caption{ Stationary in-degree probability tail under preferential
linking with attractiveness for an out-degree with $P_{out}(k)
\sim \frac{1}{k^{2+\beta}}$ as a function of
$\delta=\frac{A}{E_{o}}$ and $\beta$. The horizontal axis
corresponds to preferential linking ($A=0$). In the separatrix
curve, $\delta=\beta-1$, $P_{in}(k)\sim
\frac{log(k)}{k^{3+\delta}}=\frac{log(k)}{k^{2+\beta}}$. }
 \label{figfases}
\end{figure}

Now, we study the relationship between the tails of the in-degree
and the out-degree distributions. In the case $A=0$, if the
out-degree distribution has finite expectation ($\langle D_{out}
\rangle <\infty$) and a scale invariant tail,
$P_{out}(k)\sim~k^{-(2+\beta)}$, it is not difficult (from
eq.~\ref{pref4} (b)) to see that the limit degree distribution and
the in-degree distribution have the following tail behavior:
\begin{equation}\label{prin}
P(k)\sim P_{in}(k)\sim
\begin{cases}
k^{-(2+\beta)} \quad \mbox{$0< \beta <1$}\\
log(k)k^{-3} \quad \mbox{$\beta=1$}\\
k^{-3} \quad \mbox{$\beta >1$}\\
\end{cases}
\end{equation}
Eq.~\ref{prin} constitute our second main result: if the
out-degree distribution has finite variance and a scale invariant
tail, $P_{out}(k)\sim k^{-(2+\beta)}$, then the limit in-degree
distribution has also a scale invariant tail, $P_{in}(k)\sim
k^{-\alpha}$. Moreover, for $0<\beta <1$, $\alpha$ is equal to
the out-degree exponent. This last result can explain why in so
many real networks the in and out power exponents are so similar,
taking values in a range from 2 to 3. In the case $\beta>1$,
$\alpha=3$, regardless of the value of $\beta$. For the frontier
case (finite/infinite variance) of $\beta=1$, the limit
distribution decays at a slower rate than $k^{-3}$. Precisely, it
decays as $P_{in}(k)\sim log(k)k^{-3}$. In the general case of
preferential linking with attractiveness for $P_{out}(k)\sim
k^{-(2+\beta)}$, the regimes are similar to the
non-attractiveness case.  In this case the only difference is
that there is now a separatrix curve between them, as it is shown
in
 Fig.~\ref{figfases}. The behavior is regulated by
$\delta\equiv A/E_o$ and $\beta$. For $\delta>1+\beta$ the limit
out degree $P_{in}(k)\sim k^{-(2+\beta)}$, and in this case the
(in) degree distribution has exactly the same tail as the
out-degree, even for large $\beta$. For $\delta<1+\beta$,
$P_{in}(k)$ behaves as $k^{-(3+\delta)}$.
 Finally on the separatrix curve, $\delta=1+\beta$, the behavior
is given by $log(k) k^{-(3+\delta)}$. Note that $\delta$
($A/\langle D_{out} \rangle$) can not be smaller than -1, since
$\langle D_{out} \rangle$ must be (see eq.~\ref{b}) greater than
-A.

For out-degree distributions with exponential tails, as a
geometric, Poisson, or finite range distributions, the in-degree
distribution satisfies that $P_{in}(k)\sim k^{-(3+\delta)}$, even
for negatives values of $\delta$. In~\cite{net:lambio} they show
that for the PRL citation network the out-degree distribution has
an exponential decay, and the in-degree one has a power law tail
with  $\alpha$ near 3, just as described before for the null
attractiveness case.
 We remark the following:
 a) if the model is adequate for describing a real growing
 network, and this network has an out-degree distribution with exponential
tail, and a scale invariant in-degree distribution with a power
between 2 and 3, then attractiveness parameter must be negative,
and b) if the empirical in-degree distribution has a scale
invariant tail with a power less than 2, then the model presented
here is not adequate for describing this network.  Keeping in
mind the last point, the new estimations~\cite{serrano} of the
in-degree power exponent of the WWW network, would rule out the
model for describing this particular network.

\subsection{Application: scientific publications network }
  The scientific publications network has two advantages that define it as the most ``pure'': 1)
extremely few double arrows, and 2) all the variability in the
number of out-links is ``intrinsic''. These two features guarantee
that our model (see Fig.~\ref{model}) is adequate for describing
the scientific network. Nevertheless, it is not clear which is
the attachment law ($\pi$) such that we can obtain a good mimic
of the growing network process.

\begin{figure}
\begin{center}
\includegraphics[height=0.4\textwidth]{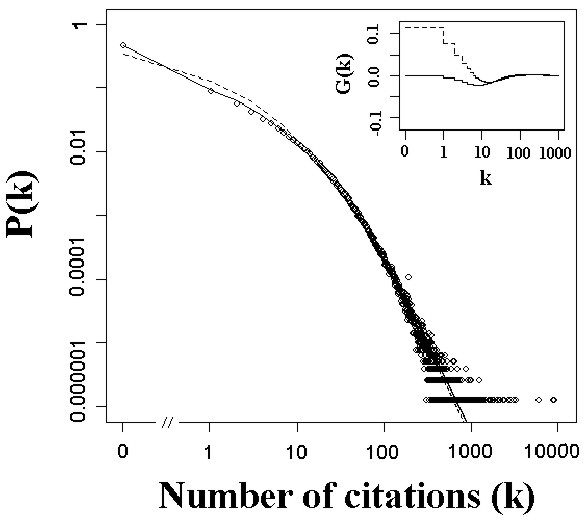}%{fig2_new.eps}
\end{center}
\caption{Citation distribution for all papers published in 1981
(from the ISI) cited between 1981 and 1997. The theoretical
citation (in-degree) curves are calculated by eq.~\ref{pref4} (c)
assuming that A=0, and the out-degree distribution is geometric,
$P_{out}(k)=p(1-p)^k$ for $k\in N_o$. The dashed line correspond
to $p=0.104$ ($T=0.115$), and the solid one to $p=0.0817$
($T=0.023$) but with $P_{out}(0)=0.3$ and
$P_{out}(k)=0.7622781p(1-p)^k$ for $k\in N$. Inset: Difference
between the empirical cumulative distribution
and the theoretical cumulative distribution. Data from~\cite{net:pagredner}.}\label{citas}% decia(3.825) (1-0.3)/(1-po)=0.7622781
\end{figure}

Fig.~\ref{citas} shows the citation distribution for all
 scientific publications published in 1981 from the ISI dataset cited between
 1981 and 1997 (see~\cite{net:redner}). Clearly, this distribution represent
 the in-degree one (see Appendix D).
  Unfortunately the out-degree distribution ($P_{out}(k)$), i.e. the number of references that
  has a randomly selected paper, has not been reported. This makes
 impossible to test the growing model by a plug-in approach (see eq.~\ref{pref4}
 (c)).
% (neither to calculate the statistical measures presented in the Section 2.2).
 Nevertheless, we take the following strategy: we suppose a geometric
  out-degree distribution $P_{out}(k)=p(1-p)^k$ with $k\in N_o$,
  a preferential linking on degree attachment law (eq.~\ref{pref2} with $A=0$), and finally
  we estimate $p$.
   Probably the empirical out-degree distribution ($P_{out}(k)$) does not fall in any
  family of parametric distributions. However, a well estimated in-degree
 distribution will be a positive result,
   since the in-degree distribution is obtained as a result of a theoretical computation
   based on the out-degree distribution. In order to estimate $p$,
   we first compute the average number of citations in the ISI network ($\langle cites \rangle=8.573$)
    and impose the condition that $\langle cites \rangle=\langle
references\rangle=\overset{\infty}{\underset{k=0}{\sum}}k
   P_{out}(k)=8.573$ we obtain  $p=1/(9.573)$. The dashed line in
   Fig.~\ref{citas} correspond to this case.
   If we estimate separately the case $k=0$, and assume that the out-degree distribution is such that
    $P_{out}(0)=a$, $P_{out}(k)=cp(1-p)^k$ for $k\in N$ with $c=(1-a)/(1-p)$,
  we obtain $p=(1-a)/8.573$ after taking the mean value condition.
    Curiously, for $a=0.3$ ($p=0.0817$) the theoretical in-degree probability (solid line)
    is extremely similar to the empirical one
   in all the range of the distribution, which can not be achieved with an oversimplified
   model where $P_{out}(k)=\delta_{k=m}$.
   This is not the only $P_{out}(k)$ that fits perfectly well, hence we do not assert that
   the estimated $P_{out}(k)$ must be similar
   to the real cites distribution. Moreover, the estimated $P_{out}(k)$ does not seem very adequate,
   since under this probability distribution 30\% of all scientific publications do not contain
any reference (yet, note that in~\cite{net:price} it was reported
that 10\% of all publications do not contain any reference).

In order to have a better notion of the goodness of fitness we
compute the Kolmogorov statistic,
\begin{equation}\label{asombr}
T=\underset{k \in N_0}{max} |G(k)|=\underset{k \in N_0}{max}
|F_{\widehat{P}_{in}}(k)-F_{P^{theo}_{in}}(k)|
\end{equation}
where $F_{P}(k)$ is the cumulative distribution,
$F_{P}(k)=\overset{k}{\underset{j=0}{\sum}}P(j)$, $P^{theo}_{in}$
correspond to the theoretical in-degree distribution showed in
eq.~\ref{pref4} (c) assuming a particular $P_{out}(k)$, and
$\widehat{P}_{in}$ correspond to the empirical citation
distribution. One advantage of the proposed estimator in
eq.~\ref{asombr} is that it is possible to test whether the model
(including the attachment law) is adequate for describing the
real network. In our application, the null hypothesis is $H_o$:
the real growing network has an underlying link attachment law
that is preferential on degree.
 For the simplest case where $T$ compares an empirical distribution with a theoretical one, but without
 estimating parameters, the null hypothesis will be rejected (at a 0.05 level of significance) only
if $T>0.0015$.  In the case shown with solid line $T=0.023$, and
for the case  where $P_{out}(k)$ is geometric (dashed line)
$T=0.115$. Clearly, $T$ is a good measure for ranking models (or model selection). The
inset of Fig.~\ref{citas} shows the function $G(k)$ for both
out-degree distributions proposed, for the geometric (dashed
line) case the maximum distance between the cumulative
distributions (see eq.~\ref{asombr}) occurs at $k=0$, and for the
other case (solid line) at $k=10$.

As we mentioned at the beginning of this section, the model is
adequate for the scientific publication network, but the
attachment law is completely unknown. We have proposed one,
preferential linking on degree, but we do not have the possibility
 to corroborate it. This is one of the reasons why we are
going to study the model under different attachment laws. The
only weak argument in favor of the law given by eq.~\ref{law1}, is
that review papers, that have a huge number of references, are
typically highly cited compared with regular articles that have a
small number of references. In this way, the correlation between
$D_{in}$ and $D_{out}$ will be positive, which is a virtue of the
law defined in eq.~\ref{law1}.

\subsection{Different attachment laws}
Clearly, it may happen that for a real network the informal
checks (covariance, variance and conditional expectation)
discussed before might be not consistent with the observables of
the model. In this case, three things may be happening: 1) the
link attachment law is not adequate, 2) the model is not correct,
or 3) both before. The first point is related to the mechanism of
linking: preferential, uniform, non linear preferential, or may
have some age dependency as described
in~\cite{net:mendes2,net:kjara}.  The second point correspond to
the growing mechanism, that can be seen as the core of the model.
For example, updating of nodes, or a very high proportion of
double links can be present, that are not considered in the
model. In this section we discuss only the alternative where the
attachment law is different from the one proposed in
eq.~\ref{law1} (preferential linking on degree), but the core of
the model remains true.

\subsubsection{Preferential linking on in-degree}
In~\cite{net:mendes} they studied a model where the attachment law
depends on the in-degree and on the attractiveness. The proposed
law was the following:
\begin{equation}\label{mend}
\pi_{in}^k=\frac{(A+k)N_{in}^k}{\underset{j \in
N}{\sum}(A+j)N_{in}^j},
\end{equation}
where $N_{in}^k$ is the number of nodes with in-degree equal $k$.
 In principle, this can be a good law for the scientific
 publications network.
The joint attachment law in this case is given by:
\begin{equation}\label{pp}
\pi_{deg,out}^{j,k}=\frac{j-k+A}{\langle D_{out}
\rangle+A}P_{deg,out}(j,k),
\end{equation}
where we have used that $\langle
D_{in}\rangle=\overset{\infty}{\underset{k=0}{\sum}}kP_{in}(k)=\langle
D_{out}\rangle$.
 Replacing eq.~\ref{pp} in eq.~\ref{din_gral2}, it is very easy to compute the stationary probabilities:
\begin{equation}\label{attrac2}
\begin{aligned}
%& P_{out}(k)=p_k \quad \quad \quad \quad  \quad \quad \quad \quad \quad \quad \quad \quad \quad \quad \quad \quad  (a)\\
& P_{in}(k)=\frac{\Psi(k+A,2+\delta)}{\Psi(A,1+\delta)} \quad \quad \quad \quad \quad \quad \quad \quad  \quad \quad \   (a)\\
& P(k)=\frac{1}{\Psi(A,1+\delta)}
\overset{k}{\underset{j=0}{\sum}}P_{out}(j) \Psi(k-j+A,2+\delta) \quad (b)\\
& P_{in,out}(j,k)=P_{deg,out}(j+k,k)=P_{in}(j)P_{out}(k)\quad  (c) \\
\end{aligned}
\end{equation}
where $k, j \in \mathbb{N}_o$. This case is specially easy to
solve because, for a randomly selected node,
 the number of out-links ($D_{out}$) and the number of in-links
 ($D_{in}$) are independent random variables ($P_{in,out}(k,j)=P_{in}(k)P_{out}(j)$). This mean:
 \begin{equation}\label{rmen}
 \begin{aligned}
& r=0   \quad \quad \quad \quad  \quad \quad \quad \quad
\quad \quad \quad  (a) \\
& \langle D_{in}/ D_{out}=k\rangle=\langle D_{out}\rangle \quad\quad (b)\\
& \langle D_{out}/ D_{in}=k\rangle=\langle D_{out}\rangle \quad \quad (c). \\
 \end{aligned}
 \end{equation}
One big difference between the previous attachment law
(eq.~\ref{law1}) and this one (eq.~\ref{mend}) is that
$P_{in}(k)$ depends only on the mean number
 of out-links ($\langle D_{out}\rangle$) by $\delta$ ($\delta=A/ \langle
 D_{out}\rangle$), and not on the shape of the out-degree distribution ($P_{out}(k)$).
  For $A>0$ and $k>>1$,  $P_{in}(k)$
behaves as $k^{-(2+\delta)}$ no matter which is $P_{out}(k)$
(only depends on $\langle D_{out}\rangle$).
 Therefore, under the attachment law given by eq.~\ref{mend},
 the tail of the out-degree distribution does not give any information
 about the tail of in-degree distribution, contrary to what happens for the law of
 eq.~\ref{law1}. In addition, for this new attachment law the correlation between $D_{in}$ and $D_{out}$
 is zero (eq.~\ref{rmen} (a)), and the conditional expectation of
 $D_{in}$ ($D_{out}$) given $D_{out}=k$ ($D_{in}=k$) does  not
 depend on $k$ (eq.~\ref{rmen} (b) and (c)).

 Note that $\pi^k_{in}$ in eq.~\ref{mend} is well defined only for positive or zero values of
 attractiveness. But, only strictly
 positive values of $A$ are interesting,
  since % GIVEN THAT
  for $A=0$ we get that the stationary probability is $P_{in}(k)=\delta_{k=0}$.
  This last result is easy is to understand: new nodes appear but they can
  not be pointed by other nodes ($A=0$), and in this way
the network will be formed by almost all nodes with zero in-links
and only a few (given by the initial condition of the network) with many
in-links. Clearly, in the limit  $n\rightarrow \infty$ the
proportion of nodes with $k$ in-links goes to a delta function
($\delta_{k=0}$).

\subsubsection{Uniform attachment law}
It is thus clear that even when preferential linking is an
accepted mechanism of link attachment, it is necessary to
study~\cite{net:tipos_attach,net:kapp} alternative types. For the
uniform attachment law on degree:
\begin{equation}
\begin{aligned}
& \pi^k=\frac{N^k}{\underset{j \in N}{\sum}N^j} \\ &
\pi_{deg,out}^{n,k}=P_{deg,out}(n,k) \\
\end{aligned}
\end{equation}
%\begin{equation}
%\pi_{deg,out}^{n,k}=P_{deg,out}((n,k))
%\end{equation}
by means of the same technology (replacing $\pi_{deg,out}^{n,k}$
in eq.~\ref{din_gral2}) we obtain:
\begin{equation}
\begin{aligned}
&P(k)=\frac{1}{1+\langle D_{out}\rangle}
\overset{k}{\underset{j=0}{\sum}}P_{out}(j)
(\frac{\langle D_{out}\rangle}{1+\langle D_{out}\rangle})^{k-j}\\
&P_{in}(k)=\frac{1}{1+\langle D_{out}\rangle}(\frac{\langle D_{out}\rangle}{1+\langle D_{out}\rangle})^{k}\\
\end{aligned}
\end{equation}
Note that, $P_{in}(k)$ depends only on $\langle D_{out}\rangle$
(and not on $P_{out}(k)$), and decays exponentially fast. For an
out-degree with $P_{out}(k)\sim k^{-(2+\beta)}$, $P(k)$ behaves
as $k^{-(2+\beta)}f(k)^{-1}$, where $f(k)$ is an increasing
function of $k$ that grows more slowly than $log(k)$. It is
important to remark that for empirical (finite) networks, the
$f(k)^{-1}$ term will be very difficult to discriminate ($f(k)$
grows at a rate slower than $log(log(k))$). This behavior may be
hard to ``separate'' from $P(k)\sim k^{-(2+\beta)}$, but the
in-degree distribution will sort out any possible confusion about
the link attachment law.

\section{Conclusions}
For the model presented here, we showed a simple way to compute
the stationary probabilities. This model was constructed in order
to take into account the main features of real directed growing
networks with the property that almost all the variability in the
number of out-links is ``intrinsic'' (see Section 2).
 From the stationary Property, we showed how to compute
 the stationary joint in-out
  degree distribution for an arbitrary out degree distribution, and arbitrary link attachment law
  ($\pi$). We studied three different $\pi$'s, paying special attention
  to the
  preferential linking on degree with attractiveness mechanism ($\pi_{n}^k=
\frac{(A+k)N_{n}^k}{\underset{j \in \mathbb{N}}{\sum}
(A+j)N_{n}^j}$).
    Once obtained the joint probability, we compute:
    \begin{enumerate}
\item[(1)] $P_{in}(k)$ as a
function of $P_{out}(k)$.
\item[(2)] The correlation between $D_{in}$ and
$D_{out}$.
\item[(3)]  The conditional expectation of
$D_{in}$($D_{out}$) given $D_{out}$($D_{in}$).
\end{enumerate}
 From $P_{in}(k)$ we studied the
 relationship between the distribution tails, giving a possible
explanation for the in/out degree tail relationship
reported for many real networks. The statistical measures mentioned in (2) and (3) were
 studied for the WWW network, obtaining a good
agreement with some of the analytical results presented in this
paper. Nevertheless, we cannot say that the model is appropriate
to describe this network (an important part of the variability
would be not ``intrinsic'').

Finally, we showed an application to the scientific publications
network. In this network:
\begin{enumerate}
\item[(a)] New publications continuously~\footnote{ 
Probably a non-homogeneous Poisson process provides
a good description of the arrival of new publications. But as we are interested in
asymptotic distributions, which are independent (except
in the pathological cases where explosions might occur) of the arrival process,
it is sufficient to study the time step process, where in each
step a new publication is aggregatted.} appear (growing network) and do not disappear.

\item[(b)] The structure is rigid. Published papers cannot change their references, only new papers can change the number of citations of already published works.
% between the interarrival time nothing in the network is modified,

\item[(c)] The publication that is forthcoming has a non predictable number of references, $D_{out}$ (random variable)

\item[(d)] Even knowing $D_{out}$, the cited papers by the forthcoming publication are unpredictable (there is a law of attachment, $\pi$).
\end{enumerate}
The model we proposed considers the four points mentioned above.
 The main difference with other models, is that
 the number of out-links (references)
of a new node (paper) is treated now as a random variable. Therefore, if the
distribution of the number of references ($P_{out}(k)$) is known,
an important part ((a),(b) and (c)) of the scientific network will be well
described by the model. But, the distribution of the number of references of the forthcoming publication
(out-degree distribution) has not been reported. In addition, the attachment law ((d)) of the scientific
publication network is completely unknown, and difficult to estimate it.
 Thus, we proposed a simple out degree distribution (geometric) and an
attachment law of preferential linking on degree (we also
consider  preferential linking on in-degree and uniform
attachment). With these two assumptions, we found a very good fit.
 This application also served to discuss how
to compare various models. In this matter, we proposed a measure (eq.~\ref{asombr})
frequently used in statistics to compare two distributions.

% The remaining part ((d)) of the scientific
%publication network, is the attachment law. This is completely unknown,
%and difficult to estimate it.
 % We do not have strong evidence supporting the
%proposed law, only we have an obvious suspicion that reviews
%(have large $D_{out}$) are highly cited compared with regular
%articles (small $D_{out}$), being this consistent with the
%proposed $\pi$.

From a modeling point of view, we see our results as a further
step from which more complex models may be built in order to be
closer to reality. The model can be seen as the skeleton to
construct more sophisticated models. For example, it does not
seem difficult to incorporate in the model double links (a mixed
out-links distribution) in order to be closer to the metabolic
network, or some updates in the nodes to mimic the WWW network.
Other important issue to explore is what happens when $P_{out}(k)$
depends on time in a simple parametric way. This last point is
related with accelerating networks~\cite{net:accel}.

\vspace{1cm}
%\begin{acknowledgement}
\textit{We thank A. Calabrese, A. Cuevas, M. O'Connell, and G. Solovey for
critical reading of the manuscript,  I. Armend\'ariz, and P.
Ferrari for useful discussions, and A.L. Barab\'asi  and S.
Redner for their generosity in sharing network data. % Comments of
%anonymous referees are most appreciated.
}%\end{acknowledgement}
\begin{appendix}

\section{Comments on the model}
Being rigorous, the model as it was presented in Section 2 is not
well defined. Yet, as we discuss in this appendix, this is not a
serious problem (all the results presented before hold).
 The difficulty is that $P_{out}(k)$ is any
probability distribution. In particular, it includes the ones that
take infinitely values (such as geometric, or any one with
exponential or power law tails). The problem can be stated as
follows: if a new node, for example has 1000 links and the
network has 100 nodes, ¿what do we must do with the remaining 900
links?.

We describe below the correct form of the model (that can be
implemented):
\begin{enumerate}
\item[(1)] Initially the network consists of $n$ nodes connected in a
given arbitrary way.
\item[(2)] At each time step starting from $n+1$, say time step $m$, a node with $\widetilde{D}^{m}_{out}$
outgoing-edges appear.  $\widetilde{D}^{m}_{out}$ is a
random variable with law $Q_{out}^m(k)$ ($Q_{out}^m(k)\equiv
P(\widetilde{D}^{m}_{out}=k)$, and $\underset{k\in
\mathbb{N}}\sum P(\widetilde{D}^{m}_{out}=k)=1$).
\item[(3)] Each new directed edge points out to an existing node with some
probability law $\pi_{m}$ (uniform, preferential linking, etc.).
\end{enumerate}
The distribution of the number of out-links from a new node at
time $m$ (the networks has $m-1$ nodes) is defined by the
following equation:
\begin{equation}\label{deftr}
  Q_{out}^m(k)=P(D_{out}=k/D_{out}< m).
\end{equation}
$Q_{out}^m(k)$ is the conditional distribution of $D_{out}$ given
$D_{out}\leq m-1$. From definition~\ref{deftr} is very easy to see
that $Q^m_{out}(k)$ converge to $P_{out}(k)$,
% the
%previous definition satisfies that as the network grows
%$Q^n_{out}(k)$ converge to $P_{out}(k)$,
\begin{equation}
\underset{m\rightarrow \infty }{lim}Q_{out}^m(k)=P_{out}(k),
\end{equation}
as the network grows, where $P_{out}(k)$ is the distribution
defined a priori (see Section 2). %From this last convergence we
%can see
%have that %, it is very important to remark that for
%for the model with this correction (we have only changed
%$P_{out}(k)$ by $Q_{out}^n(k)$) the same asymptotic results that
%were obtained for model presented in Section 2 holds. The general
%conclusion would be:``small effects disappear at $\infty$''.  One
%example of this, was explained in the Section 2.4.1 were we
%discuss why for A=0, $P_{in}(k)$ converges to $\delta_{k=0}$.
From this last convergence we can see
 that the model with this correction (we have only changed
$P_{out}(k)$ by $Q_{out}^m(k)$) has exactly the same asymptotic
behavior that was %were
obtained for the model presented in Section 2.
% Therefore, the
%asymptotic results presented in Sections 2.1 and 2.2 hold for this
%corrected model.
Therefore, all the results presented in this paper also hold for
the corrected model. The general conclusion would be:``small
effects disappear at $\infty$''. See, for instance Section 2.4.1
were we discuss why for A=0, $P_{in}(k)$ converges to
$\delta_{k=0}$.
% Therefore, the
%asymptotic results for the corrected model are equal to the
%results presented in Sections 2.1 and 2.2.
% are equal to the results presented in
%Sections 2.1 and 2.2.
%\section{Appendix}
\section{A closed equation for $P(k)$}
If we were only interested on the stationary degree distribution
($P(k)$), the computation is much easier than the one presented
in Section 2.1, since there is a closed equation for $P(k)$. The
growing network dynamics is given by:
\begin{equation}\label{din2}
\begin{aligned}
& N_{n+1}^k=N_{n}^k+\Delta^k_n \quad \quad \quad \quad \quad \quad \quad \quad \quad (a) \\
&\Delta^k_n=\delta_{D_{out}=k}+\overset{D_{out}}{\underset{i=1}{\sum}}\delta_{Y_i=k-1}-\delta_{Y_i=k}
\quad (b)\\
\end{aligned}
\end{equation}
where $\{Y_i\}_{1\leq k \leq n}$ is a sequence of
independent and identical distributed random variables,
taking value $k$ ($k\in \mathbb{N}$) with probability $\pi_{n+1}^k$.

\textbf{Property:}  $\vec{P}\equiv(P(1),P(2),\dots, P(k),\dots)$
is the solution of:
\begin{equation}\label{proper2}
\langle \Delta^k_n / \frac{\vec{N}_{n}}{\underset{k \in
\mathbb{N}}{\sum}N_{n}^k}=\vec{P}\rangle=P(k)
%\rangle\frac{N_{n+1}^k}{\underset{k \in
%\mathbb{N}}{\sum}N_{n+1}^k}/\frac{\vec{N}_{n}}{\underset{k \in
%\mathbb{N}}{\sum}N_{n}^k}=\vec{P}\rangle=P(k)
\quad \quad \forall k \in\mathbb{N}.
\end{equation}
Replacing $\Delta^k_n$ by eq.~\ref{din2} (b) in
eq.~\ref{proper2}, we get:
\begin{equation}
\langle
\delta_{D_{out}=k}+\overset{D_{out}}{\underset{i=1}{\sum}}\delta_{Y_i=k-1}-\delta_{Y_i=k}/\frac{\vec{N}_{n}}{\underset{k
\in \mathbb{N}}{\sum}N_{n}^k}=\vec{P}\rangle=P(k).
\end{equation}
 From this last equation
it is trivial to obtain that the stationary degree probability
satisfies:
\begin{equation}\label{equil}
P(k)=P_{out}(k)+(\pi^{k-1}-\pi^k)\langle D_{out}\rangle
\end{equation}
where $\pi^k$ is the stationary probability that a new link is
attached to a node with degree $j$. Under preferential linking on
degree linking with attractiveness, the stationary  attachment
law, $\pi^k$,  remains equal to $\frac{(k+A)P(k)}{\langle D
\rangle+A}$.  Replacing $\pi^k$ in eq.~\ref{equil}, and using
$\langle D \rangle=2\langle D_{out} \rangle$, it is easy to
conclude that the limit degree distribution ($P(k)$) is given by
eq.~\ref{degree2}.
\begin{equation}\label{degree2}
P(k)=\Psi(k+A,3+\delta)
\overset{k}{\underset{j=1}{\sum}}\frac{P_{out}(j)}{\Psi(j+A,2+\delta)}.
\end{equation}

%In order to be clearer, we describe first a sketch of correct
%algorithm of the model that do not have the problem mentioned
%bove, and from this we show that the model presented as in the
%paper will give the same stationary probabilities that the one
%obtained by the algorithm. Let
%We can not guarantee that the cumpotation is correct since we are
%...
\section{WWW network }
\begin{figure}
\begin{center}
\includegraphics[height=0.6\textwidth]{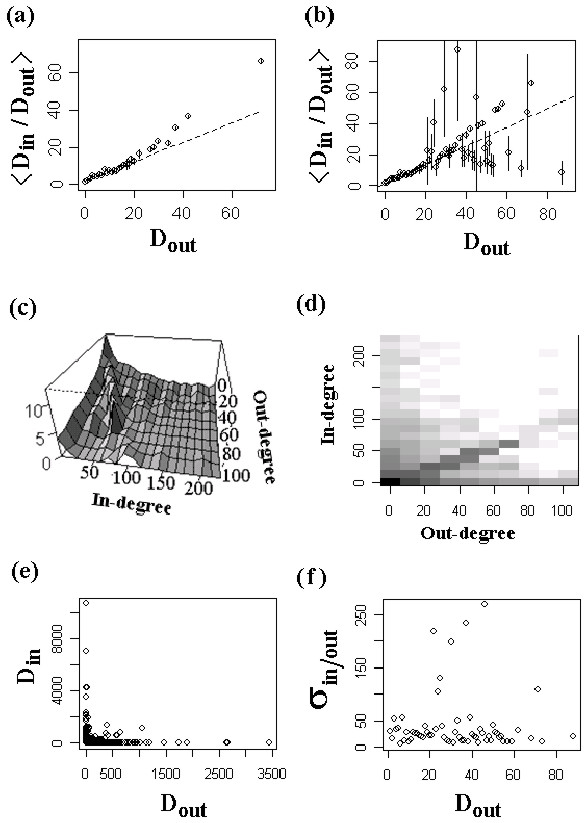}%{fig2_new.eps}
%\begin{figure}
%\begin{center}
\end{center}
\caption{Conditional mean of  $D_{in}$ given  $D_{out}$, when for
each value of $D_{out}$ there exist at least: (a) 500, and (b) 30
nodes. Data presented as a confidence interval of 95\%. (c) and
(d) Different representations of the joint in-out density of the
links in a node. (e) Scatter plot of $D_{in}$ as a function of
$D_{out}$. (f) Conditional standard deviation of $D_{in}$ given
$D_{out}$, $\sigma_{in \diagup out}$.
 }\label{barabasi}% decia(3.825) (1-0.3)/(1-po)=0.7622781
\end{figure}

%\subsubsection{WWW network}
As we have mentioned in the Section 2.2.1, it is difficult to find
 articles on networks that report
 the simple descriptive measures (covariance,
 variance and conditional expectation) for nodes discussed here.
 However, a detailed statistical analysis  of the topological
properties of  four different WWW networks have been reported
recently~\cite{serrano}. In~\cite{serrano} the covariance and the
variance of the number of out-going links ($D_{out}$) and in-going
links ($D_{in}$) are reported, which we give in Table 1.
%have been
%reported~\cite{serrano} for four different WWW networks.
%Table 1
\begin{table}[htb]
\begin{center}
\begin{tabular}{c|cccc}
  & $Cov(D_{in},D_{out})$   &  $Var(D_{out})$  & $Var(D_{in})$ \\% & $\Delta$ \\
\hline
WBGC01 & 155.682   &  171.61  & 40080.04 \\ % & 0.097 \\ %
WGUK02 & 524.244 & 750.76 & 20534.89  \\ %0.355 \\
WBGC03 & 348.486   &  870.25  &  54980742 \\ % & 0.856 \\
WGIT04 & 3478.75 &   4502.41    & 776866 \\ % & 0.257 \\
\end{tabular}
\caption{Descriptive statistical measures for 4 WWW networks.
Data from~\cite{serrano}.}
\end{center}
\end{table}
The first thing that can be noted is that for all the domains
studied  $Var(D_{out}) < Var(D_{in}) $, consistent with
eq.~\ref{cor2}. Moreover, $Cov(D_{in},D_{out})$ and
$Var(D_{out})$ have similar values (consistent with
eq.~\ref{cov}), the relative differences seems large only for
WBGC03.
%In order to improve the notation, we define the right term of
%eq.~\ref{cor},  the \underline{fraction} between the standard
%deviations, as R.
In order to compare in a better way these last two quantities,
Table 2 shows $r$ and $R\equiv
\sqrt{\frac{Var(D_{out})}{Var(D_{in})}}$ for the same data. We
can see that WBGC01 and WGIT04 have very similar values of $r$
and $R$ (see eq.~\ref{cor}).
\begin{table}[htb]
\begin{center}
\begin{tabular}{c|cc}
  & $r$   &  $R$   \\
\hline
WBGC01 & 0.0594   &  0.0654   \\
WGUK02 &  0.1335 & 0.1912  \\
WBGC03 &  0.0016   &  0.004  \\
WGIT04 & 0.0588 &   0.0761  \\
\end{tabular}
\caption{Correlation (r) and R for 4 WWW networks. Data computed
from Table 1.}
\end{center}
\end{table}
 In order to study the relationship between $\langle D_{in}/ D_{out} \rangle$
 and $D_{out}$ is necessary to have the complete data. At this point, we analyze the
 WWW data obtained from~\cite{net:pagbara} presented in~\cite{net:bara3}.
  We built up a database with the information of the
 number of out-links and in-links ($(D_{out},D_{in})$) for each of the 325729 nodes.
 In order to have a good estimation of the conditional expectation, we first restrict
the study to the values of $D_{out}$ such that there exist at
least 500 nodes. % (the complete graph can be found in the Appendix C).
 Fig.~\ref{barabasi} (a) shows the relationship between $D_{out}$ and the conditional mean of
 $D_{in}$ ($\langle D_{in}/ D_{out} \rangle$) given $D_{out}$. Interestingly, there is a strong relationship
between both. For values of the $D_{out}$ smaller than 20 there is
a clear linear relationship  between
 them. A robust regression (least median of squares) estimation between
$\langle D_{in}/ D_{out} \rangle$ and $D_{out}$ gives a slope of
0.523 and an intercept of 1.739. In the case $D_{out}$ is greater
than 20 it seems that $\langle D_{in}/ D_{out} \rangle$ grows
faster than linear, but it is not clear if this effect is real
(based on Fig.~\ref{barabasi} (b)). The graph presented in
Fig.~\ref{barabasi} (b) is similar to the one in (a), but now we
study the values of $D_{out}$ such that there exist at least 30
nodes. A plot of two different representations of the joint
in-out distribution is given in Fig~\ref{barabasi} (c) and (d),
to have an idea of the shape of the joint law, while (e) shows a
scatter plot on a larger grid.  Besides, the in-degree variance
($Var(D_{in})=1346.85$) is greater than the out-degree one
($Var(D_{out})=461.25$), consistent with eq.~\ref{cor2}.
Fig~\ref{barabasi} (f) shows the conditional standard deviation of
$D_{in}$ given $D_{out}$,
$\sigma_{in/out}=\sqrt{Var(D_{in}/D_{out})}$. Unlike the
conditional expectation,  the conditional variance does not seem
to have any  relationship with $D_{out}$.

\begin{figure}
\begin{center}
\includegraphics[height=0.22\textwidth]{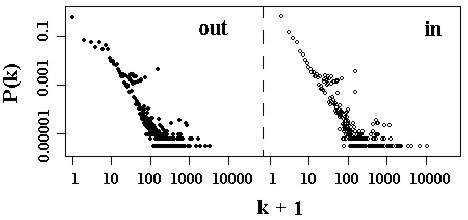}%{fig2_new.eps}
\end{center}
\caption{ $P_{out}(k)$ and $P_{in}(k)$ as a function of k+1. This graph was presented in~\cite{net:bara3}.}\label{barabasi3}% decia(3.825) (1-0.3)/(1-po)=0.7622781
\end{figure}

In~\cite{net:bara3} the authors showed the empirical out degree
($P_{out}(k)$) and in degree ($P_{in}(k)$) distributions (see
Fig.~\ref{barabasi3}), and reported a power exponent of 2.45 for
out-degree and of 2.1 for the in-degree~\footnote{Our estimations
of the exponents have some differences from the ones
in~\cite{net:bara3}, but the difference between the in and out
exponents is still appreciable.}. This is the first empirical
evidence that the model presented here can not describe in a good
way the WWW network, in the model the power law exponents are
equal. The second evidence is that $r$ and $R$ are not similar,
$r=0.2244$ and $R=0.5852$.

%Perhaps, some small modifications on the model can do it
%\underline{acceptable} for having a good mimic of the real
%growing process.

% $Var(D_{in})=1346.85,
%Var(D_{out})= 461.25$, and $R=0.5852$

%Quizás el modelo con algunas pequeñas modificaciones pueda
%describirla.

% If we believe, with the empirical evidence presented here,
% that the model presented in Fig~\ref{model},
% with  preferential linking on degree attachment law,  is \underline{adequate} for describing
% the WWW network, it is possible to estimate the attractiveness.
%   An slope of approximately 0.5 correspond to a value of
%   attractiveness 4 (using Fig~\ref{esperanza} (b) and the fact that the WWW network has $\langle D_{out} \rangle$ =4.6).
% Unfortunately, with $A=4$ $P_{in}(k)$ of eq.~\ref{pref4} (c) do not have a good much , being this perhaps a rasgo de que el modelo no es adecuado para esta red.

\section{Comment on the scientific publication network}
In the scientific publication network it is implicit that we are
under the hypothesis that the citation distribution for all papers
published in 1981 can be treated as the stationary in-degree
distribution of a growing network model. But, ¿why can be treated
in this way only studying the papers of a particular year (1981)?.
This is just because: if the total scientific network has arrived
(today in 2007) to a proportion of papers with $k$ citations that
do not change with time (stationary), then the articles published
in 1981 are a sample of this distribution.

\end{appendix}

\end{document}